# A Text Extraction-Based Smart Knowledge Graph Composition for Integrating Lessons Learned during the Microchip Design

Hasan Abu Rasheed[1], Christian Weber[1], Johannes Zenkert[1], Peter Czerner[2], Roland Krumm[2], Madjid Fathi[1]

[1] University of Siegen, Siegen, Germany
[2] Elmos Semiconductor AG, Dortmund, Germany
`hasan.abu.rasheed@uni-siegen.de`

**Abstract.** The production of microchips is a complex and thus well documented process. Therefore, available textual data about the production can be overwhelming in terms of quantity. This affects the visibility and retrieval of a certain piece of information when it is most needed. In this paper, we propose a dynamic approach to interlink the information extracted from multisource production-relevant documents through the creation of a knowledge graph. This graph is constructed in order to support searchability and enhance user's access to large-scale production information. Text mining methods are firstly utilized to extract data from multiple documentation sources. Document relations are then mined and extracted for the composition of the knowledge graph. Graph search functionality is then supported with a recommendation use-case to enhance users' access to information that is related to the initial documents. The proposed approach is tailored to and tested on microchip design-relevant documents. It enhances the visibility and findability of previous design-failure-cases during the process of a new chip design.

**Keywords:** Knowledge Graphs, Text Mining, Graph-Based Search, Lessons Learned, Microchip Design.

## 1 Introduction

Manufacturing processes produce a considerable amount of data, which contains, but is not limited to: measurements, parameters, reports and documentations [1]. With time, accessing especially the accumulated textual data becomes a long and rather inefficient search procedure, if it is approached through traditional searching methods, such as a key-word-based search. In this context, searching the available data is a mean of enhancing its visibility, aligned to the content of the respective search. However, it is not the only method of accessing data. Data visibility, to the users, can be handled through searching, hints or recommendations, among other methods. Depending on the context, one or more methods can be utilized to enable not only accessing textual data, but also extracting the specific intention of documents, such as previous experiences in the production process, in the form of lessons-learned documents from the past. For example, failures in previous production processes are usually documented along the production line. However, it is not likely that a new worker will search for all possible failures that can take place when adjusting a certain parameter in a machine. In this case, a system that automatically highlights a previous failure case, which is related to this particular parameter or this machine, has the potential to save time or prevent possible damage.

In this paper, we handle this challenge in a limited environment, which is the microchip design process. In the design phase of a new semiconductor product, the ability of the design engineer to have an overview on the database of previous design failure cases and tailored lessons learned, enables an improvement in the design process. It leads to time and material savings and quality enhancements, e.g. for the choice of combinations of electronic elements for the chip. This, in turn, is reflected as an overall quality improvement, not only of the product, but also of the design procedure itself.

Therefore, our proposed system is meant to support the design process through enhancing the visibility of design-related textual data from manufacturing-relevant datasets. It allows the design engineer to search for specific failure cases and provides automatic highlights of the previous failures that are related to the current design, which the engineer is working on. As shown in [2], especially in the field of semiconductor manufacturing, tailored instead of off-the-shelf solutions are continuously needed to enable smart solutions, despite growing amounts of relevant, tracked data.

The textual data, utilized for this approach, reveals three challenges facing the preparation and searchability:

The first challenge is the free nature of the textual input. Design failures are reported by engineers with semi-structured and unstructured formats. Free textual input may also include symbolic descriptions of elements, standard abbreviations, internal domain-specific abbreviations and, in some cases, two languages in the failure description. Those issues have to be handled, in order to prepare the available data for a search engine. Therefore, an extended text mining pipeline was developed and utilized for the preprocessing of this data.

The second challenge comes from the multisource nature of the available data. Different sources of data provide important information [3], [4], regarding design and production processes. Those sources provide different data types, which accompany the main textual one, such as numerical values or structure-related data from the design project. In order to enable the user of accessing those different sources and their corresponding data types, this work solves the challenge with the utilization of knowledge graphs. As a network of interconnected data, a knowledge graph holds the potential to represents multiple data sources in one network. Therefore, different data sources and modalities can be intelligently fused together, in form of nodes, in the graph. This allows the search engine to access multiple modalities in one search command. Moreover, information that lies within the different sources also influences the intelligent construction of the graph itself, allowing more hidden links and relations between failure cases to be discovered, and thus, enhances the overall visibility of related lessons from previous design processes.

Since the number of connections in a knowledge graph can be overwhelming, the third challenge in this work is related to the ordering and recommendation of most relevant failure cases to a certain current design. This is handled with a simple recommendation approach that utilizes both, the information gained from the text mining algorithm and the construction of the graph nodes.

In this paper, we introduce a text mining pipeline to construct knowledge graphs from product-design-focused documents with regard to the specific type of a document. Furthermore, we propose a text mining-based concept for enhancing the construction of the created knowledge graphs through the introduction of a special type of connection-focused nodes in the graph. We call this node type the linking nodes, which are proposed to enhance the connectivity between multisource and multimodal data nodes within the graph, in order to improve its searchability. The proposed approach also highlights main limitations in the application environment, which are mainly related to the data quality. The amount of data, as well as the amount of non-meaningful tokens, authoring misconceptions and irregular misspellings can influence a mining algorithm, since it can limit the amount of useful information extracted from it. We follow mitigation procedures to counteract those limitations and we address their effect on the results in the result section. The first findings of this approach are reported in this paper as follows: in the second section of the paper, the background of the utilized algorithms is presented. In the third and fourth sections, the proposed methodology is described regarding text mining and knowledge graphs respectively. The experimental results are presented in the fifth section, and the sixth section concludes this work along with its future perspectives.

## 2 Background

In order to handle textual data without an explicit pre-known structure, text mining techniques provide solutions for extracting useful information from plain unstructured texts. However, textual data can be differently structured, depending on the type of document, which may enable to extract information that is more contextualized and therefore more specific. This work will consider three types of specialized documents:
- **documents with a hierarchical structure:** which holds here information about the design patterns and potential relations between different parts of the designed chip.
- **unstructured documents:** including content that is not consistently structured.
- **documents with a column-based labeled structure:** effectively labeling textual content. In this work, this content represents structured lessons-learned tables.

To handle the multiple sourced data fusion methods are needed. Among several fusion approaches reviewed within the state-of-the-art, such as [5]–[7], intelligent semantic methods provide an important solution to integrate multiple data modalities, such as text mining results and available structural information, in one framework that is queryable and can be a base for a decision support system.

### 2.1 Text Mining

Text mining is a collection of methods and algorithms intended for information extraction and knowledge discovery from structured, semi-structured and unstructured textual data [8], [9]. In the mining process, several techniques work individually and collaboratively to process raw data and utilize it in the knowledge extraction. Such techniques and methods include pre-processing, document retrieval, classification, topic discovery and modelling, and word association and information extraction. Furthermore, the combination of outputs of individual methods can be used to further inform future models as seen in [10], in an industrial context.

In the pre-processing phase, raw textual data is cleaned from undesired strings, or strings that are without meaning for the respective case, such as special characters and email addresses. After cleaning, the text is divided into smaller pieces that can be analyzed. This is referred to as tokenization, in which the large corpus is cut into sentences, phrases or words, each called a token. In tokenization, some identifiers such as punctuations can be used to separate the individual tokens. Tokens are then also scanned to eliminate less meaningful tokens that are referred to as stop words. Stop words can be [11]: punctuations or pronouns, frequent words or, in some cases, non-frequent vocabulary. Extra steps can be performed such as stemming and lemmatization, in which different forms of the same name or verb are turned back to their basic dictionary form.

In the context of semantic representation, information extraction methods that can discover the relations between textual data points are of considerable importance. This is due to the role they play in mapping the extracted knowledge and linking its information in a queryable manner, where the user can navigate from a textual entry or a search term, reaching relevant information. Information extraction has traditional methods that are manually crafted in many cases [12]; as well as intelligent methods based on machine and especially deep learning [13], [14]. In [14], authors consider relation extraction as a classification problem, in which different predefined relation types can be used to recognize the belonging of a relation in the corpus to a certain category. In this context, probabilistic methods accompany deep learning methods to perform the classification. Such approaches have been utilized by the authors in [15] to extract relations from free textual data and an existing knowledge base. Their approach builds a domain specific knowledge graph, which can be queried in applications such as question answering and chatting. The graph is built on two levels, a static level from the predefined database, based on a domain specific terminology, and a dynamic graph that is built automatically from the free text. Relation extraction is conducted by the deep neural network through a matching task on a large training dataset. The resulting static and dynamic graphs are merged into one, allowing an effective knowledge representation and querying. Similar approaches can also be observed in [16]–[19].

With the development of search engines and the popularity they witnessed, retrieving the most relevant results for a search query got more focus in the literature. With textual data arranged in documents, several methods have been implemented for document matching, in order to find the result that best corresponds to a user input. A document in this context can be defined as a full or partial textual corpus, such as a paragraph or a sentence.

Document matching is essential in applications such as search engines, question answering and recommender systems. In such fields, the matched documents are also ranked and ordered according to their relevance to the user query [20]. Ranking can be based on a weighting criterion such as the well-known Term Frequency-Inverse Document Frequency (TF-IDF) approach, which measures the importance of frequent words in a document, in contrast to rarely existing words throughout a set of documents. The assumption in this algorithm is that a frequent word in one document can be a feature that describes the document content, which can also be the case for words that exist in only few documents within a larger dataset. This allows a more differentiated retrieval of documents when such feature words are included in the user query.

## 2.2 Knowledge Graphs

Knowledge graphs, or knowledge maps, are graphical representations of a knowledge base. Whether ontologies or other semantic representations that are based on relations between graph entities, those networks have been a robust approach utilized for knowledge representation, information extraction and decision support.

Semantic knowledge representations can enhance or be enhanced by other technologies related to data preparation, such as text mining, information extraction, deep learning and natural language processing [21]. In their summary on the utilization of knowledge graphs with text mining, authors in [21] highlight the rising role of knowledge graphs in text-centric retrieval, especially for search system applications. Their review focuses on linking graph entities, the retrieval of those entities and the role of text mining in the retrieval process. In this context, an entity of a knowledge graph is one entry in this graph, or a node, which has its attributes and relations to other nodes/entities in the graph. Since each node in the graph has normally a textual form of information attached to it, such as annotations, properties or names, linking graph nodes can then be defined as the process of identifying the relevance of this textual content between two nodes in the graph. This can then be expressed explicitly as an additional relation between them. Retrieving graph nodes, on the other hand, refers to the process of identifying and extracting nodes in the graph, which are related to a certain query of the user. In other words, it is determining the relevance of a node, based on a relevance measure between the text of a search term and the textual content of the nodes in the graph. Since, in a text extraction-based graph generation, graph nodes are representing extracted text from text documents, retrieval models are required, which are called retrieval models. A standard retrieval model combines heterogeneous and semi-structured information about the node, e.g. its attributes, into one representation that can be static, as in [22], or dynamic such as in [23]. Textual data can enhance the node retrieval through supporting the matching process between the node attributes, on one side, and the textual query on the other side. This can be accomplished on a level that uses keywords or on a complex features level, where machine learning algorithms can play an effective role in learning the matching functions [24].

Knowledge graphs can also utilize other data mining methods to enhance the search and matching tasks. In [25], authors develop a concept to integrate the Resource Description Framework (RDF), as a machine understandable format, with human understandable formats using natural language. Their approach is meant to support the representation of large heterogeneous and distributed datasets, which in turn helps in building decision support systems.

Sinoara, Antunes and Rezende in [26] review the most utilized text mining approaches in semantic descriptions of data. In their study, the authors point out the majority of algorithms that exploit text classification and clustering in support of the graph construction or the matching and search within the constructed graph. They also highlight the role of domain experts in the utilization of textual data in knowledge graphs, since they can provide valuable information about the potential relations between the graph's nodes. However, they find that domain experts are seldom integrated in the mining process, and they assign that to the complexity that is added to the system when considering interactive interfaces, which engage the expert user in the graph construction process.

Moreover, the authors in [26] describe the visualization process in the context of knowledge graphs and text mining. Several approaches and tools are mentioned, which consider recommended methods of visualizing the relations within mined textual data, such as the Pinda tool [27], which adopts a hierarchical representation of texts using the k-means clustering algorithm for grouping the mining results. In [28], approaches are introduced to extract sub-knowledge-graphs as procedural representations, to partially exploit hierarchical information in semi-structured texts.

In this paper, we utilize the results from the state of the art and propose a new concept for a document aligned knowledge graph construction, in which text mining supports introducing a new type of nodes in the graph, called the linking nodes, in order to enhance the graph's connectivity, multiple source data fusion and searchability.

## 3     Text Mining and Document Representation

In this work, textual data from multiple documentation sources has been collected for analysis. Multisource data includes maintenance reports of previous failure cases, written with a lessons-learned perspective, product descriptions and specifications, as well as design project structures and details from the semiconductor design. Since the obtained data was originally generated in a very specific context, which is the microchip design, it contains a considerable amount of symbolic descriptions, free textual inputs and remarks for the lessons-learned context, as well as semi-structured texts that include a considerable amount of misspellings, non-standard abbreviations, incomplete sentences and, in some cases, multiple languages. Handling those cases is an essential step to create a valid descriptive model of each document, which in turn will be the enabler for connecting those documents and mining the relations among them. To accomplish this goal, we utilize the previously described text mining pipeline as the first step to reduce undesired or error- containing tokens in the textual data. However, since standard text mining methods are not sufficient to create meaningful representations of the domain specific documents in hand, an extended text mining algorithm was built as a pipeline (Fig.1) that integrates an expert- and data-driven, rule-based, approach to handle domain specific differences in the textual data.

After defining the multiple formats of documents from the available dataset, all texts have been extracted from those individual documents. Then, non-symbolic special characters are removed, along with specific patterns that are automatically generated from the database's software. The remaining text is then filtered form punctuation and tokenized. The resulting tokens are lemmatized and stemmed in different phases of the procedure. Resulting tokens are also the input for the expert-supported part, where firstly, standard abbreviations are identified and isolated. Then, special symbols that are domain-specific, such as multiple underscores, are also reconstructed so that they are handled similarly between the multiple document sources.

With the reduction of non-meaningful tokens, a representative model of each document is constructed.  Document models serve the goal of representing a full document with a vector of features, which is reliable to describe the extended content of the original document. In this paper, we utilize the TF-IDF algorithm (Equ.1) to extract most descriptive tokens from a document, as well as from the overall sum of available documents. In the microchip design use case, term frequency $tf$ is capable of detecting meaningful repeating words in a design document, such as the term "oscillator" or "processor". However, this part of the TF-IDF also detects non-relevant frequent words, such as "volt", which do not hold a unique meaning in the document, although it is considerably frequent. This is why the IDF part enters the process, in order to balance the discovered terms identified in the TF part.

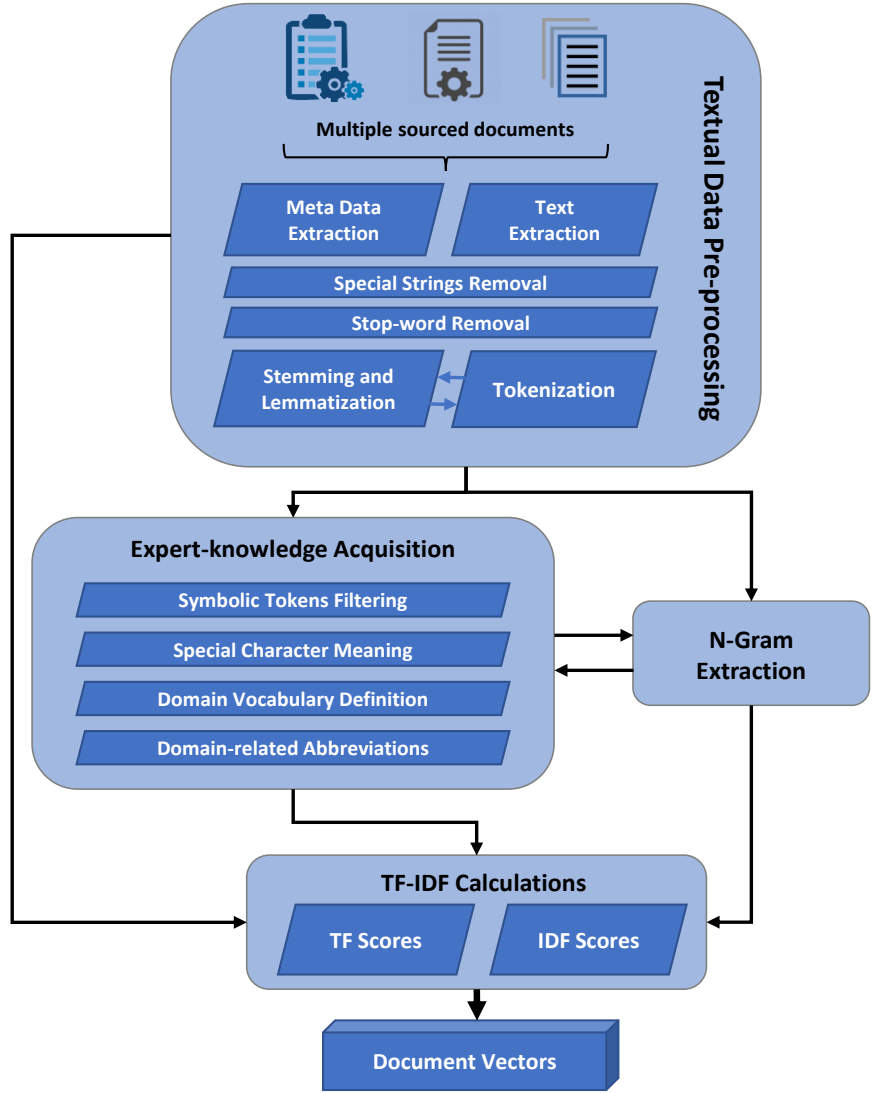

**Fig. 1** Text mining algorithm and document vector generation

The inverse document frequency $idf$ in our use case identifies the tokens that appear in a limited number of documents among the total document dataset. Based on the assumption that those terms are most likely to represent the topic of the document, they are given a high score by the IDF algorithm.

$$TF - IDF_{t,d} = tf_{t,d} \times idf_t = tf_{t,d} \times \log\left(\frac{N}{df_t}\right) \qquad (1)$$

In Equ.1, $t$ is the term in question, $d$ is the document in which the term $t$ is looked up; and $N$ is the total number of documents. Scores for the all tokens are calculated from the TF and the IDF parts. Then, a certain threshold determines the choice of tokens that will represent each document in a vector format. Document vectors provide a unified approach for describing the content of documents from the multiple sources through the most representative terms in those documents. This approach has been chosen for its benefits in the construction of the knowledge graph, as well as the fusion of information extracted from different documentation process. This, in turn, supports the search functions to perform the same query on all data sources, since they have a coherent representation in the graph. However, document vectors are not the only type of information extracted from the available data sources. Each of the available data sources contains extra information based on the source nature itself. For example, a project description, as a documentation source, inherits the information about the structure of the project, which is utilized to create links between certain documents. Moreover, failure documentations that contain free textual descriptions of maintenance procedures, also contain patterns that appear when the same person writes the case description, or when the workers in the design department use certain terminology amongst each other. Those patterns can also be mined and utilized to extract relations between maintenance documents. In order to achieve this goal, we exploit an N-gram approach in the text mining algorithm. N-grams are combinations of N number of tokens that appear in the same pattern over the textual data. In our proposed approach, we identify the largest useful number of combinations that reveal possible patterns in the failure cases' description, which is three tokens or named as trigram. Then we utilize all N-grams equal or below this threshold, i.e. unigrams, bigrams and trigrams, in order to enhance the relation extraction between documents that contain those N-grams. However, unlike the TF-IDF based vectors, the N-grams approach does not represent the document's topic based on the frequent and unique terms in it. Therefore, instead of including those N-grams in the document vectors, we utilize the structure of the constructed knowledge graph itself and introduce a new type of nodes to the graph that is mainly based on the N-gram approach. This maintains the document-specific vectors and allows the addition of new relations to the graph through the N-grams. We call those nodes the linking nodes; and they serve the goal of creating new connections between graph nodes, based on the mined possible patterns in the textual data. The following section presents a detailed description of this concept, along with the graph construction procedure.

## 4 Knowledge Graph Structure

Although obtaining data from multiple sources poses the potential of supporting the textual data with other data modalities, it also reveals a main challenge concerning the fusion of those modalities in one representation, as well as the extraction of relations between them. In this paper, a text mining-based algorithm provides the overall system with the ability to reduce the imprecision in textual data, represent the documents in meaningful token vectors and extract possible patterns in the text represented by the extracted N-grams. Document vectors hold the potential to extract relations between the documents. However, searching for a certain document through those relations is not very different from a key-word search approach, which is a limited approach in this particular use case. Therefore, to enhance the connectivity, searchability and retrieval of design documentation, a knowledge graph is utilized for its ability to represent relations between those documents based on several criteria.

### 4.1 Knowledge Graph Construction

The construction of a knowledge graph consists mainly of two steps: defining the graph's nodes and extracting the relations between them. In this paper, a graph node is defined as a representation of an individual document. This representation is mainly accomplished with the document vector. However, since available datasets contain multiple document types that require different processing, graph nodes do yield a different meaning, depending on the specific document type they represent. In other words, although all document-representing nodes in the graph contain the feature vector of the document, each type of the documents attaches different information to the node that is representing it. In this way, a graph of multiple node types, has been constructed, in order to include all useful information extracted from multiple data sources. Examples of this complexity is the structural information that accompanies the textual data extracted from project descriptions. In this case, project documents provide not only the elements used in a certain microchip design, but also the hierarchical structure of the final chip. This information is added to those nodes in the graph which represent project description nodes.

In the proposed graph, conceptually illustrated in Fig. 2, we define three types of document representing nodes:

- **Failure Case Nodes (FC):** Which represent the maintenance documentation in the form of reports of previous design failure cases. This type of nodes is the center of the search functionality in the proposed system, since enhancing the visibility of previous failure cases, and thus lessons learned, is the main objective of the developed algorithm.
- **Project Element Nodes (PE):** This type of nodes represents the documentation of semiconductor elements that are utilized for microchip design projects. Those elements are inserted in the design of the component composition. The elements have an implicit hierarchical structure, expressing a "is part of" relationship, revealing a hierarchical structure of the project, as well as element patterns that appear in several projects. Therefore, structural information accompanies the textual description in this node type, enhancing the relation extraction process in the graph.
- **Product Specification Nodes (PS):** Which are the final and most detailed documentation of the designed product. This type of documents includes the technical details of the designed chip and is organized in a human readable form, to support design engineers as well as to inform product customers.

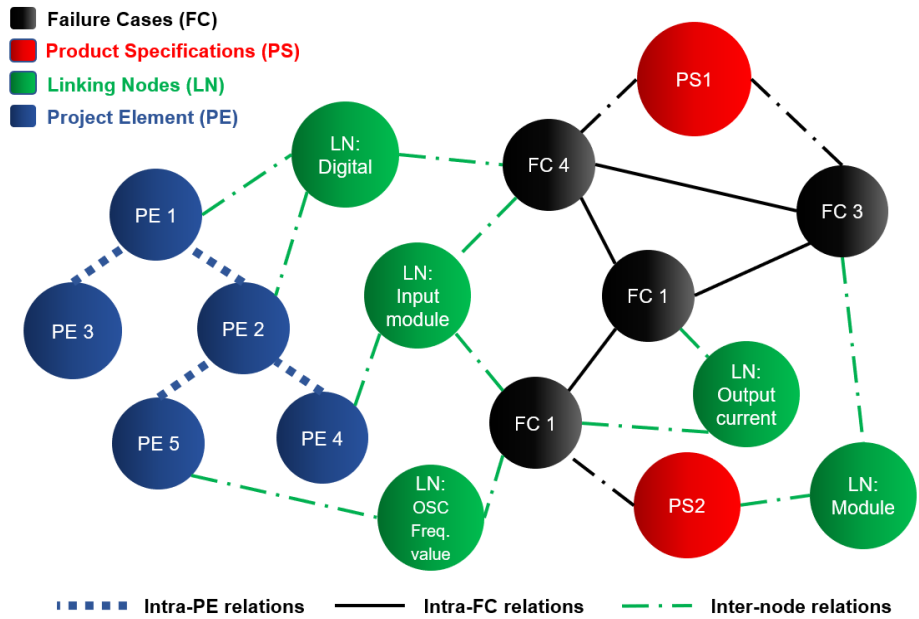

**Fig. 2** Proposed structure of the knowledge graph with node types

After defining node types, relation extraction is the following step in the construction of the proposed graph. In the proposed graph, inter-node-type and intra-node-type relations are introduced:

- **Intra-node-type Relations:** This type of relations link document nodes of the same type together. It is extracted from the specific information available for each node type. For example, intra-failure-case relations, or FC-FC relations, are extracted from the textual patterns that appear in maintenance documents, whereas intra-project-element, or PE-PE, relations reflect the hierarchical structure of the elements in the microchip design, as a parent-child relationship.
- **Inter-node-type Relations:** Which represent the relations between different types of nodes in the graph, such as FC-PS relations between failure cases and product specification documents. This type of relation is based on the shared structure of the nodes, i.e. the document vector format.

Fig. 3 shows the actual resulting knowledge graph with the distribution of three types of nodes included: FC, LN and PE nodes.

### 4.2 Linking Nodes (LN)

In order to express the relatedness of concepts, based on their co-occurrence in the text, within proposed graph, we introduce in this work a new type of nodes that introduces weighted relations to the constructed network of nodes. Relations defined by linking nodes do not only provide new connections on both inter-type and intra-type relations, but also enable a better search-result filtering and recommendation, through the potential to differentiate between node and relation weights. Inserted linking nodes in the graph are weighted by default, since they are based on the N-gram approach. This influences the weight of relations that connect them to other node types in the graph. For example, connections to trigram nodes have more weight than those to unigrams, since they denote a higher similarity between the nodes' content.

Additionally, graph's connectivity is enhanced by this approach, which can improve the visibility of failure cases, as they are then connected not only through single and direct concepts but also through indirectly related ones, revealing potential hidden relations between documents. However, results show that introducing new relations does not linearly correlate with an improvement of searchability. We have found that after a certain level of new connections introduced to the graph, search results converge to a one output where detailed changes are negligible. Table 1 in the next section gives an overview on the number of relations that are introduced by different types of nodes in the proposed graph.

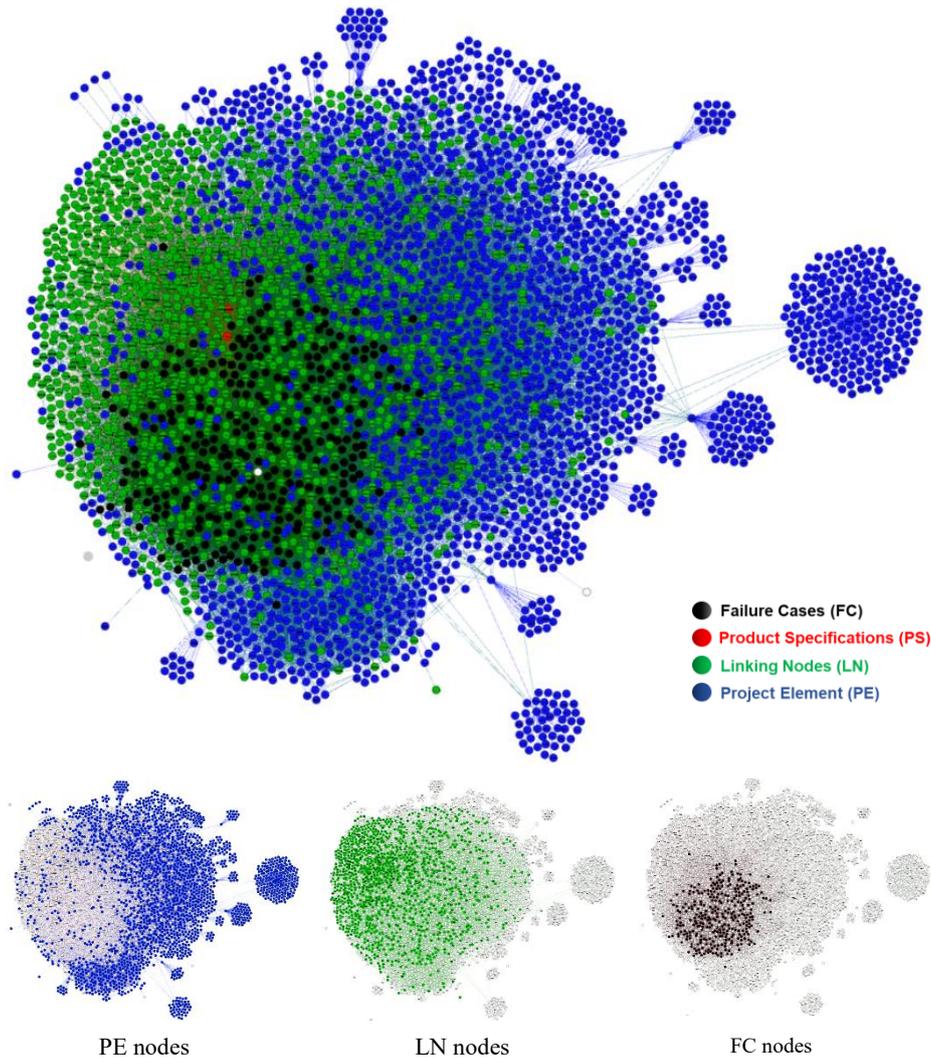

**Fig. 3** Final constructed knowledge graph

## 5 Experimental Results from Graph Search

A graph-based search is conceptualized and utilized in this paper to exploit the connectivity of graph nodes for the purpose of enhancing document visibility to the user. The experiment is set for the application domain of microchip design, utilizing design documentation and maintenance cases.
Two retrieval scenarios are handled in this paper:
1- direct search for a failure cases using search terms.
2- indirect search in the graph upon the insertion of a new element in the project design.

The latter case is meant to highlight relevant maintenance documents and failure cases, which are related to an inserted component within a specific design structure, to the design engineer during the design phase.
The knowledge graph search mechanism that is adopted in this paper includes, for both cases, two stages. Following a search input from the user, the algorithm will find matching nodes and allocates, firstly, nodes in the graph that have a direct relation to the search term node with a distance of one. We refer to those nodes as the base nodes. Relations are based upon the document vector, N-grams or the structural information. Secondly, the algorithms allocated nodes that are connected to the ones discovered in the first step. Those new allocated nodes do not relate directly to the search term, but they are connected to it with other types of relations that have been extracted in the graph construction process.
In this mechanism, the number of search results can be controlled through leveraging several parameters. We utilize two experimental parameters in this paper, which are search depth and relation weights.

Search depth is defined as the number of intermediate relations that separate a base node from the furthest node connected to it. Within this definition, if the search depth is equal to 1, search results will contain base nodes, that are directly matching the query, and those that are connected to them. The deeper the search is, i.e. the more relation steps are followed between the nodes, the higher the number of results is. In fact, the number of results increases exponentially with the increase of search depth. Therefore, other parameters or search strategies are needed to govern the number of research results provided to the user.

To handle this issue, relation weights are utilized in this work, alongside the search depth, to order search results by relevance and provide the user with a reasonable number of connected documents. In the passive search scenario, triggered by adding components to the design process, this method forms the base of a simple recommendation process, which offers the user hints about relevant maintenance documents that are related to an inserted element and, in general, the current design.

Table 1 provides an overview of the relations extracted and utilized in the construction of the knowledge graph. Both inter-node-type and intra-node-type relations are shown in the table, and they highlight the condensed interlinking between failure case node in the graph. This is due to the fact that retrieving previous failure cases, as lessons learned, is the main objective of the graph search. This is also clear from Table 2, where the percentages of node numbers, relative to the total node count in the generated graph, is reported. It is noticeable that within the existing dataset, the number of maintenance documents, represented be FC nodes, is relatively low compared to other types of documents. This is a usual limitation facing industrial datasets, due to the level of digitalization in the factory and the confidentiality of available data. However, through the text mining of maintenance reports' textual content and the thorough relation extraction between them, they still possess the highest number of links in the graph, which positively influenced their retrieval rates, reaching between 28% and 43% more visibility, depending on the search configurations.

**Table 1** Number of relations introduced by different types of nodes

| Relation types | Node types | Number of relations |
| --- | --- | --- |
| Intra-node-type | FC-FC | 11146 |
|  | PE-PE | 1862 |
| Inter-node-type | PE-LN | 9634 |
|  | FC-LN | 12988 |

**Table 2** Number of nodes in the graph

| Node type | Node count | Percentage in the graph |
| --- | --- | --- |
| Failure Cases (FC) | 304 | 9% |
| Linking Nodes (LN) | 1171 | 37% |
| Project Elements (PE) | 1862 | 54% |

Utilizing relation weights, as well as the depth of search as parameters to set the configurations of the search process, the number of search results for three search cases is illustrated in Fig.4, alongside the number of total relations in the graph, before and after inserting LN and PE relation types into the graph. For a unified search term and one level of search depth, the resulting number of maintenance documents that are visible to the user has increased by 28% when introducing the newly defined relations, which are based on the linking node and the intra-PE approaches. With the same number of relations, increasing the search depth to the second level has increased the visibility of documents by up to 43%, compared to the original case with only FC relation.

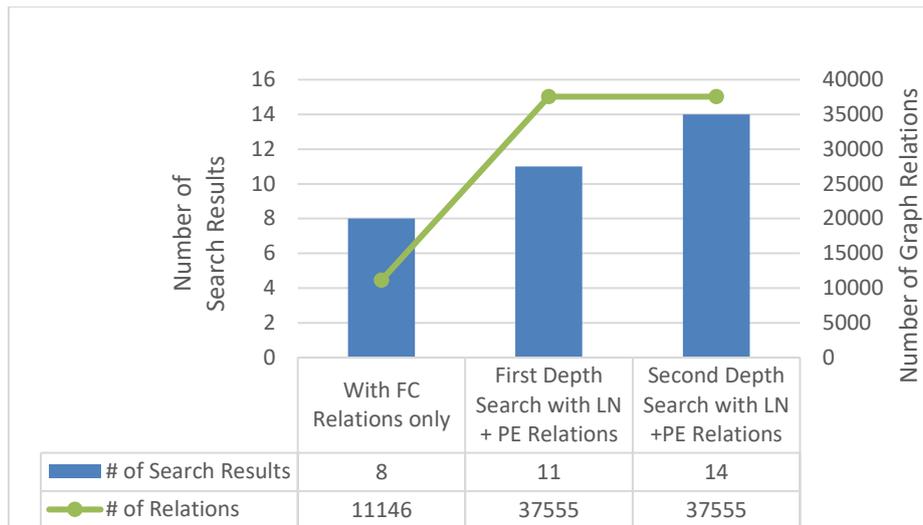

**Fig. 4** Effect of LNs and structural relations on the search results and graph connectivity

## 6 Conclusion and Future Perspectives

In this paper, a new approach for a document tailored construction of knowledge graphs has been proposed, based on an extended text mining pipeline and the introduction of a linking node type, in order to enhance the graph's utilization through connecting and weighting similar concepts. The objective of the research is to improve the microchip design quality, through increasing document visibility and findability for design engineers, based on the constructed graph. Multisource, multimodal, data sets have been utilized in the definition of document models and the extraction of interrelations amongst them. An extended text mining algorithm was utilized and tailored to define and construct the different types of extraction-based nodes in the knowledge graph. Nodes and relations were built upon the information extracted from multiple data types. An approach to enhance the graph's construction and inter-linking has been proposed and tested in order to highlight the increased visibility of relevant queried design documentations to the user, through a search and recommendation scenario. Results of this approach show that the introduction of linking nodes positively influenced the document visibility and access by up to 43%.

The presented approach is specifically tailored to the individual types of documents present in the given use-case, which are utilized in a way to maximize the extraction and interlinking of relevant information, e.g. through further exploit their explicit, known structure. However, the exploited structural and contextual information is equally present in other failure, best-practice and design related documentations of products in other domains of manufacturing and can thus be applied and tailored to more domains as a novel integrated approach of knowledge graph generation and utilization.

The next steps of this research are to include the development of extended recommendation algorithms that build upon the proposed relation-weight method, in order to improve the user feedback from the system, as well as the relevance of the search results.

**Acknowledgment:** This work was supported by the EU project iDev40. This project has received funding from the ECSEL Joint Undertaking (JU) under grant agreement No 783163. The JU receives support from the European Union's Horizon 2020 research and innovation program and Austria, Germany, Belgium, Italy, Spain, Romania.